\documentclass[MSNbibl,nameyear,dvips]{arxstspdf}
\usepackage{flushend}
\usepackage{stfloats}

\volume{27}
\issue{2}
\pubyear{2012}
\firstpage{197}
\lastpage{198}
\doi{10.1214/12-STS376D} 
\referstodoi{10.1214/11-STS376}



\begin{document}
\begin{frontmatter}
\vspace*{6pt}
\title{Discussion of ``Statistical Modeling of Spatial Extremes'' by
A. C. Davison, S.~A.~Padoan and M. Ribatet}
\runtitle{Discussion}

\begin{aug}
\author[a]{\fnms{Benjamin} \snm{Shaby}\corref{}\ead[label=e1]{bshaby@stat.berkeley.edu}}
\and
\author[b]{\fnms{Brian J.} \snm{Reich}\ead[label=e2]{brian\_reich@ncsu.edu}}
\runauthor{Benjamin Shaby and Brian J. Reich}

\address[a]{Benjamin A. Shaby is Postdoctoral Researcher, Department of Statistics, University of
California, Berkeley, USA \printead{e1}.}
\address[b]{Brian J. Reich is Assistant Professor,
Department of Statistics, North Carolina State University \printead{e2}.}

\end{aug}



\end{frontmatter}

The review paper on spatial extremes by Davison, Padoan and Ribatet is
a most
welcome contribution. The authors cover quite a lot of ground, making
connections between different approaches while highlighting important
differences. In particular, we applaud their careful attention to model
checking, which can be difficult in general but particularly so for spatial
extreme value models.

\section{Prediction and Model Validation}

With its extensive set of diagnostics for evaluating model fit, this
paper provides
a nice template for practitioners to follow. However, perhaps the most important
feature of spatial models is their ability to predict at unobserved
sites. The
account presented here does not address the prediction problem, which
is both a
critical task in its own right and a tool for comparing models. More traditional
spatial analyses typically include various performance metrics to evaluate
prediction at a withheld test set of observation locations. While spatial
prediction is difficult for the max-stable process models described in the
paper, computational tools to accomplish this task do exist (\cite{wang-2011a}).
Spatial prediction for copula models is considerably more straightforward.

However, we note that even with predictions at hold-out locations in hand,
evaluating model skill at reproducing extremal quantities requires some care.
Clearly, the metrics used in traditional geostatistical analysis such
as mean
squared prediction error are unsatisfying for block-maximum data.
Rather, we
recommend the quantile score and the Brier score for threshold
exceedences, as
discussed and justified by \citet{Gneiting-2007}. These metrics are specifically
tailored to evaluate the tail of the predictive distribution and
therefore seem
more appropriate in this context.


\section{Toward Hierarchical Bayesian Max-Stable Models}

In their discussions of the relative merits of various approaches, the
authors highlight the ability of hierarchical Bayesian models to
represent richly flexible structures for underlying marginal
parameters. As they point out, however, the conditional independence
assumption made in the Bayesian analyses they discuss hamstrings the
model's ability to produce spatial association in process realizations.
Indeed, others have also shown that failing to properly account for
spatial dependence can lead to dramatic underestimation of uncertainty,
and thus undercoverage of posterior intervals, for the GEV parameters
and return levels (\cite{fuentes-2010a}).

The authors correspondingly laud the ability of max-stable processes to capture
joint behavior across spatial locations, but lament the restriction to
relatively simple underlying structures that pairwise likelihood
fitting of
max-stable process models imposes. The trade-off between flexible marginal
modeling and realistic spatial dependence modeling is almost treated as an
inherent conundrum, almost analogous to a Heisenberg's uncertainty
principle for
spatial extremes. But we want to have it both ways!

The authors rightly note that the unavailability of joint likelihoods for
max-stable process models appears to render their inclusion in hierarchical
Bayes\-ian models problematic. We view surmounting this obstacle as a welcome
challenge! As they note, prog\-ress has already been made. For example,
\citet{ribatet-2012a} specifies such a hierarchical model, but
replaces the
joint likelihood with a~pairwise likelihood and modifies the resultant\break MCMC
sampler using an asymptotic argument. The resultant sample from the
``posterior'' distribution appears to have desirable frequentist
properties.\break
While this approach may not be completely satisfying in that it is
computationally intensive and it does not pass the Bayesian purity
test, it
represents nice out-of-the-box thinking and is a clear step in the right
direction.

The authors also mention our recent manuscript, which describes an
approach to
hierarchical max-stable process modeling that we really like because it is
fully Bayesian, straightforwardly produces predictions at unobserved locations,
and can be fit to large data sets. Our approach suffers a bit because
it does
not readily generalize to most of the max-stable process models
mentioned here
by the authors. While possibilities certainly exist to expand on our approach,
and efforts are underway to do just that, we expect that completely
novel angles
and insights will be brought to bear on the rapidly-evolving field of Bayesian
analysis for spatial extremes.

\section{Spatial Modeling of High Quantiles}

Finally, we note that while the authors focus exclusively on the asymptotic
extreme value thoery, other approaches do exist. Asymptotic arguments and
parametric assumptions are clearly needed to estimate very extreme quantities,
such as the 10,000 year return level required by the Dutch Delta Commission.
However, there are many important applications where the focus is less extreme.
For example, one may be interested in determining the effect of climate change
on the 10-year return level of daily maximum temperature. Given the
vast amount
of meteorological data collected in the past century, there may be sufficient
data to justify a less restrictive and more robust approach such as quantile
regression. Classical quantile regression (Koenker\break (\citeyear{Koenker-2005})) gives a
nonparametric estimate of covariate effects on a quantile (equivalent
to a
return level) of interest. Recently, semiparametric Bayesian quantile regression
models have been proposed, including methods for spatial data
(\cite{Lum-2010}; \cite{Reich-2011}; \cite{Reich-2012}; \cite{Tokdar-2011}). An advantage of the
Bayesian approach to quantile regression is the possibility of
centering the
prior of a flexible quantile model on a parametric extreme value distribution,
and thus hopefully exploiting asymptotic arguments as the data deem appropriate.
Comparing, and ideally merging, quantile regression with extreme value analysis
may be a~promising line of future research.

%

\end{document}